\documentclass{optica-article}

\journal{opticajournal} 

\articletype{Research Article}

\usepackage[capitalize]{cleveref}
\usepackage{soul}
\crefname{section}{Sec.}{Secs.}
\Crefname{section}{Section}{Sections}

\crefrangelabelformat{section}{#3#1#4--#5\crefstripprefix{#1}{#2}#6}
\crefrangelabelformat{figure}{#3#1#4--#5\crefstripprefix{#1}{#2}#6}
\crefrangelabelformat{equation}{(#3#1#4--#5\crefstripprefix{#1}{#2}#6)}

\crefmultiformat{equation}%
{\edef\crefstripprefixinfo{#1}Eqs.~(#2#1#3}%
{,#2\crefstripprefix{\crefstripprefixinfo}{#1}#3)}%
{,#2\crefstripprefix{\crefstripprefixinfo}{#1}#3}%
{,#2\crefstripprefix{\crefstripprefixinfo}{#1}#3)}

\usepackage{lineno}
\usepackage{verbatim}
\def\sF{\mathcal{F}}
\def\sP{\mathcal{P}}
\def\sR{\mathcal{R}}
\def\sT{\mathcal{T}}
\DeclareMathOperator{\Tr}{Tr}
\newcommand{\PL}[1]{\textcolor{teal}{{\em PL:} {#1}}}
\newcommand{\jml}[1]{\textcolor{cyan}{{\em JML:} {#1}}}
\newcommand{\rd}[1]{\textcolor{red}{#1}}
\newcommand{\SamSu}[1]{\textcolor{violet}{{\em SS:} {#1}}}
\begin{document}

\title{High-resolution tunable frequency beamsplitter enabled by an integrated silicon pulse shaper}

\author{Chen-You Su,\authormark{1,2} Kaiyi Wu,\authormark{1,*} Lucas M. Cohen,\authormark{1,3} Saleha Fatema,\authormark{1} Navin B. Lingaraju,\authormark{4}  Hsuan-Hao Lu,\authormark{5} Andrew M. Weiner, \authormark{1,$\ddagger$}  Joseph M. Lukens,\authormark{1,5} and Jason D. McKinney\authormark{1,*}}

\address{\authormark{1}Elmore Family School of Electrical and Computer Engineering and Purdue Quantum Science and Engineering Institute, Purdue University, West Lafayette, IN 47907, USA
\\\authormark{2}Department of Physics and Astronomy, Purdue University, West Lafayette, IN 47907, USA
\\\authormark{3}Photonic and Phononic Microsystems, Sandia National Laboratories, Albuquerque, NM 87123, USA
\\\authormark{4}The Johns Hopkins University Applied Physics Laboratory, Laurel, MD 20723, USA 
\\\authormark{5}Quantum Information Science Section, Computational Sciences and Engineering Division, Oak Ridge National Laboratory, Oak Ridge, TN 37831, USA

\email{\authormark{*}wu1871@purdue.edu, mckinnjd@purdue.edu}
~ {$\ddagger$}Deceased.
}




\begin{abstract*} 
We demonstrate high-fidelity, tunable, and ultrafine-resolution on-chip frequency beamsplitters 
using a quantum frequency processor based on an integrated pulse shaper with six spectral channels. Near-ideal Hadamard gate performance is achieved, with fidelity $\sF > 0.9995$ and modified success probability $\widetilde{\sP} > 0.9621$ maintained across  frequency spacings from 2--5~GHz and down to as few as four spectral pulse shaper channels. The system's support of frequency spacings as narrow as 2 GHz significantly surpasses prior bulk demonstrations and enables arbitrary splitting ratios via spectral phase or modulation index control. These results establish a scalable and resource-efficient platform for integrated frequency-bin quantum photonics, opening new directions in quantum information processing---including densely parallel single-qubit operations and multidimensional gate implementations.
\end{abstract*}

\section{Introduction}
Photons are highly attractive carriers of quantum information due to their transportability in both free space and optical fiber, resilience to decoherence, and compatibility with both room-temperature and cryogenic environments. In addition, they support high-dimensional quantum information processing across a wide range of physical degrees of freedom (DoFs), including path, orbital angular momentum, and time-frequency encodings \cite{flamini2018photonic, erhard2020advances}. Among these DoFs, frequency-bin encoding, which encodes quantum information in discrete frequency modes~\cite{lukens2016frequency, kues2019quantum, lu2023frequency}, is particularly well suited for quantum communication and networking~\cite{Tagliavacche2025, KhodadadKashi2025, Pang2025}. It is naturally compatible with existing telecommunications infrastructure: different frequency components copropagate within standard single-mode fibers, sharing a common spatial mode that provides intrinsic stability against environmental perturbations, and can be efficiently demultiplexed using dense wavelength-division multiplexing components. 

However, photons of different frequencies are orthogonal eigenmodes and do not naturally interact with one another in linear materials. Coherent manipulation across frequency modes relies on second- or higher-order wave mixing in nonlinear optical media. While progress in this area has been substantial, most existing techniques require strong optical pump fields ~\cite{kobayashi2016frequency,kobayashi2017mach,mcguinness2010quantum,clemmen2016ramsey,Joshi2020}, which introduce pump-induced noise photons and necessitate filtering or cryogenic suppression. Scaling such systems to higher dimensions further demands multiple pump fields and tailored phase-matching~\cite{Oliver2025}. As an alternative route, the quantum frequency processor (QFP) paradigm~\cite{lukens2016frequency} offers a pump-free, intrinsically low-noise approach that is well suited for high-dimensional scaling. Leveraging a fixed set of components---two electro-optic phase modulators (EOPMs) and a pulse shaper---experimental QFPs have enabled a variety of high-fidelity operations, including single-qubit unitaries ~\cite{ lu2018electro, lu2020fully}, two-qubit entangling gates~\cite{Lu2019a}, high-dimensional discrete Fourier transforms~\cite{ lu2022high}, and Bell-state analyzers~\cite{Lingaraju2022}. These results establish the QFP as a promising platform for advanced frequency-encoded quantum information processing.

To date, QFPs have only been realized using off-the-shelf, tabletop components. These systems are typically limited to frequency-bin spacings above 18 GHz \cite{lu2018electro} due to the resolution constraints of commercial pulse shapers, which are based on free-space diffractive optics.
Unlocking the full potential of high-dimensional quantum state manipulation requires finer spectral resolution: on one hand, it increases the number of accessible frequency modes within a fixed optical bandwidth; on the other hand, it relaxes the bandwidth requirements of the waveform synthesizer and the modulation-speed demands of the electro-optic modulator, thereby enabling more sophisticated temporal waveforms capable of coherently mixing modes separated by multiple bin spacings~\cite{lu2022high,myilswamy2025chip,kues2017chip}.
Beyond performance considerations, the achievement of true scalability in terms of size, weight, and power (SWAP) necessitates the transition of QFPs to chip-scale implementations. Integrated photonics offers the potential to address both scaling issues \emph{simultaneously}---i.e., finer resolution and smaller footprint---underscoring the importance of photonic integration for the future of frequency encoding.


In this work, we demonstrate a significant advance toward fully integrated QFPs~\cite{nussbaum2022design} by replacing the bulky, tabletop pulse shaper traditionally used in QFPs with a compact, silicon-based six-channel integrated device occupying less than 1 mm$^{2}$. Building on the promising, albeit relatively sparse, history of microring-based on-chip spectral shapers~\cite{agarwal2006fully,Khan2010,wang2015reconfigurable}, 
our specific foundry-fabricated circuit~\cite{cohen2024silicon, wu2025chip} achieves ultrafine spectral resolution for frequency-bin operations with markedly reduced bin spacings. Validating performance through a tunable Hadamard gate with high fidelity and success probability, 
we confirm consistent gate performance for frequency-bin spacings ranging from 2--5 GHz, the minimum spacing marking a ninefold improvement in spectral resolution over prior tabletop implementations \cite{lu2018electro}. 
Furthermore, tests of spectrally parsimonious versions of the gate filling only four total bins 
point to the potential for ultratight parallelization.  
Finally, we demonstrate control over the splitting ratio, either by tuning the phase applied through the pulse shaper or adjusting the modulation index of the EOPMs.
Overall, our demonstration of high-quality, narrowly spaced, and tunable frequency beamsplitting marks a significant milestone toward scalable, fully on-chip frequency-bin quantum state manipulation.

\section{Background}
\begin{figure}[tb!]
\centering\includegraphics[width=13.2cm]{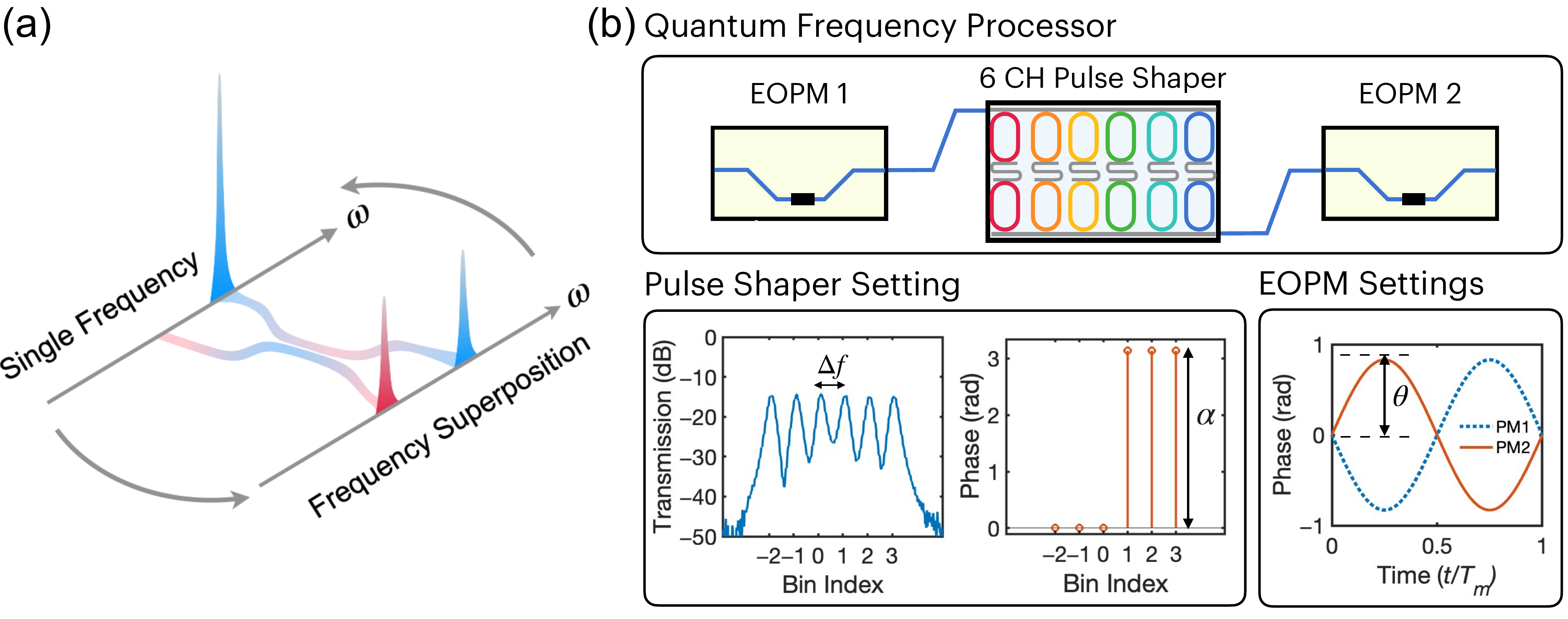}
\caption{(a) Conceptual illustration of a frequency beamsplitter, the spectral-domain analogue of a spatial beam splitter. (b) Schematic of the QFP (top) and its spectro-temporal configuration (bottom). The pulse shaper is programmed with six transmission channels spaced by $\Delta f\equiv \Delta\omega/2\pi$, with spectral phase settings $(0,0,0,\alpha,\alpha,\alpha)$. The two EOPMs are driven by out-of-phase RF signals with a common modulation index $\theta$. 
}
\label{fig1}
\end{figure}

In analogy to spatial beamsplitters, which split photons into or mix them from two lightpaths, frequency beamsplitters operate in the spectral domain [Fig.~\ref{fig1}(a)] by coherently transforming a single-frequency photon into a superposition of two frequency modes, or vice versa. In our implementation, the frequency beamsplitter is realized using a QFP composed of an on-chip pulse shaper interleaved between two EOPMs. The first EOPM generates frequency sidebands, enabling mode mixing but also introducing undesired higher-order sidebands outside the computational basis. Subsequent pulse shapers and EOPMs can then return these sidebands to the original space through interference~\cite{lukens2016frequency,lu2023frequency}. For the three-element frequency beamsplitter of interest here, this compensation obtains when the second EOPM operates exactly out of phase with respect to first, 
with the pulse shaper between them imparting spectral phases controlling the mixing weight or frequency-bin ``reflectivity''~\cite{lu2018electro,lu2018quantum, Lu2023a}. 

Mathematically, the QFP can be written as a linear transformation $V_{mn}$ between the input and output annihilation operators $\hat{a}_n$ and $\hat{b}_m$, each associated with a specific frequency bin $\omega_n=\omega_0+n\Delta\omega$ and $\omega_m=\omega_0+m\Delta \omega$, respectively:
\begin{equation}
\label{eq:V}
\hat{b}_m = \sum_{n=0}^{N-1} V_{mn} \hat{a}_n,
\end{equation}
where $N$ is the number of bins in the computational space ($N=2$ for a $2\times 2$ beamsplitter). For the configuration illustrated in Fig.~\ref{fig1}(b), we can write the full transformation of the QFP as (see Appendix~\ref{appendixA} for details)
\begin{equation}
V_{mn} =\sum_{k=1}^{B/2} \left[ J_{m+k-1}(\theta)J_{n+k-1}(\theta) + e^{i\alpha} J_{m-k}(\theta)J_{n-k}(\theta) \right],
\label{eq:Vbs}
\end{equation}
where $J_\ell(\cdot)$ denotes the Bessel function of first kind. This expression is defined for $n\in\{0,...,N-1\}$ and $m\in\mathbb{Z}$; i.e., the inputs are limited to the $N$-dimensional computational space, while the outputs can in principle appear in any bin.
The theory encompassed by \cref{eq:Vbs} fully describes all tunable frequency  beamsplitters explored below, each of which is uniquely specified by the triple $(B,\alpha,\theta$), where $B$ is the channel number of the pulse shaper (assumed even), $\alpha$ the applied phase shift, and $\theta$ the EOPM modulation index.

For convenience in comparing to an ideal target unitary matrix $U$, we define the $N\times N$ computational-basis submatrix $W$ with elements $W_{mn}= V_{mn}$ for $m,n\in\{0,...,N-1\}$; then the gate fidelity $\sF$ follows as~\cite{lukens2016frequency}
\begin{equation}
\label{eq:F}
\sF  = \frac{|\Tr U^\dagger W|^2}{\Tr U^\dagger U\Tr W^\dagger W} = \frac{1}{N}\frac{|\Tr U^\dagger W|^2}{\Tr W^\dagger W},
\end{equation}
which quantifies the similarity between the experimentally obtained transformation and the target unitary matrix $U$. We next introduce the modified success probability  $\widetilde{\sP}$
\begin{equation}
\label{eq:modP}
\widetilde{\sP}  = \frac{\Tr W^\dagger W}{\Tr V^\dagger V}
\end{equation}
to quantify the fraction of output power remaining in the computational basis, averaged over all input states. Notably, in contrast to bulk pulse-shaper-based QFPs where the success probability is typically defined with a normalization factor $\Tr U^\dagger U$ in the denominator \cite{lukens2016frequency}, we adopt this modified definition of success probability throughout the main text to accommodate the limited number of spectral channels available in the on-chip pulse shaper. This adjustment is discussed in detail in Appendix \ref{appendixB}.

In this work, we focus on the specific case of a two-dimensional beamsplitter ($N = 2$), for which we define frequency-bin reflectivities and transmissivities quantifying the fraction of mode conversion or preservation, respectively:
\begin{equation}
\label{eq:RandT}
\sR_{0\rightarrow1} \equiv |W_{10}|^2\ ; \quad 
\sR_{1\rightarrow 0}\equiv |W_{01}|^2\ ; \quad 
\sT_{0\rightarrow0} \equiv |W_{00}|^2\ ; \quad 
\sT_{1\rightarrow1} \equiv |W_{11}|^2. 
\end{equation}
For the parameter set $(B, \alpha, \theta) = (6, \pi, 0.8283)$ drawn from prior work~\cite{Lu2019b,nussbaum2022design}, direct evaluation of \cref{eq:Vbs,eq:F} gives $\widetilde{\sP} = 0.9747$ and $\sF = 0.999999$ relative to the ideal Hadamard gate $U=H=\frac{1}{\sqrt{2}}\left(\begin{smallmatrix} 1 & 1 \\ 1 & -1\end{smallmatrix}\right)$. In practice, precise measurement of the modulation index beyond the third decimal place is difficult. We therefore initially adjust the modulation index to approximately $\theta \approx 0.83$~rad and then fine-tune the RF power driving the EOPMs to achieve the most balanced beam splitting results.


\section{Experimental setup}\label{section3}

\begin{figure}[tb!]
\centering\includegraphics[width=12cm]{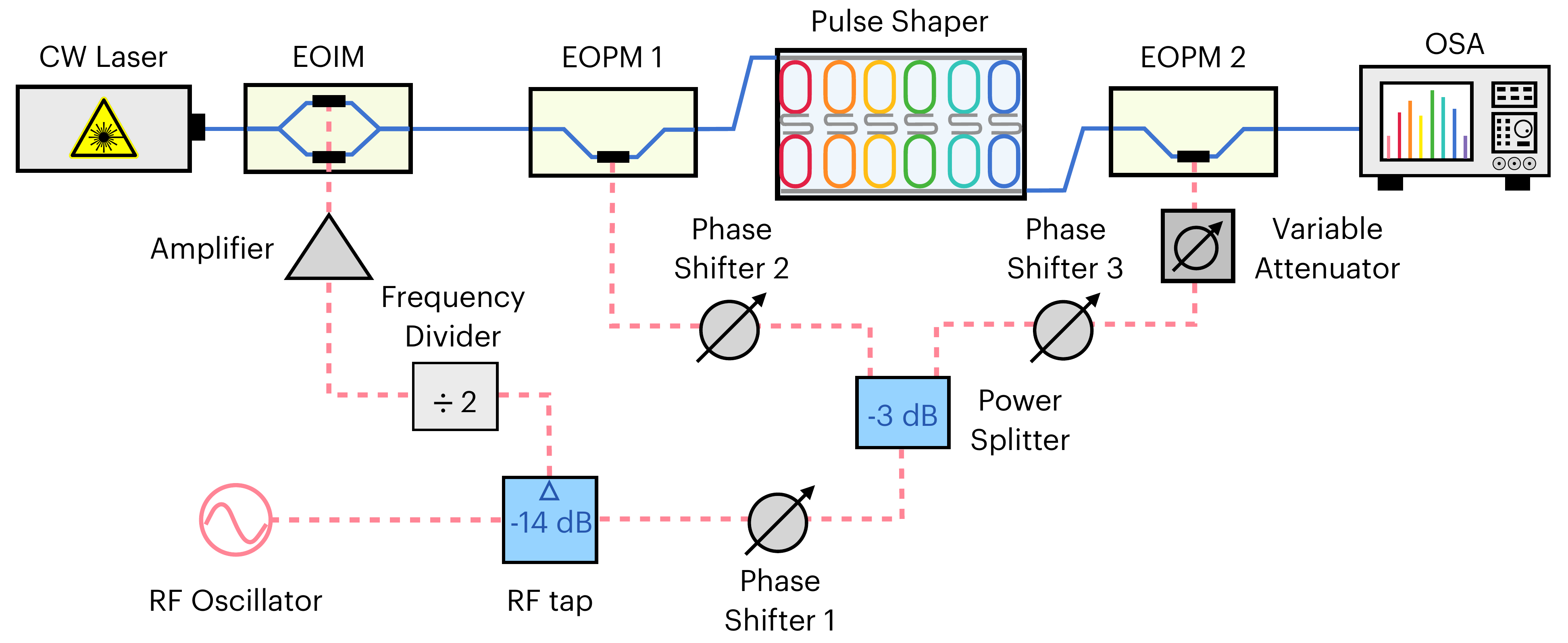}
\caption{Experimental setup. EOIM: Electo-optic intensity modulator; EOPM: Electo-optic phase modulator; OSA: optical spectrum analyzer. Solid blue lines represent the optical path, while dashed pink lines indicate the RF path. The EOIM is activated only during dual-line testing. Phase Shifter 1 can be used to adjust the relative phase between the two input spectral lines. The QFP is composed of two EOPMs interleaved with an on-chip pulse shaper. A variable RF attenuator balances the modulation indices of the two EOPMs. Shifters 2 and 3 are tuned to ensure out-of-phase modulation between EOPM 1 and EOPM 2. The output spectrum is recorded by a high-resolution OSA.}
\label{fig2}
\end{figure}


The experimental setup is illustrated in Fig.~\ref{fig2}. Our QFP configuration employs the same on-chip pulse shaper previously demonstrated in \cite{cohen2024silicon,wu2025chip}. The pulse shaper comprises six pairs (twelve in total) of racetrack resonators, each with a drop port full-width at half-maximum (FWHM) of 1.3 GHz. When the two resonators in each pair are tuned to the same central frequency---using integrated thermal phase shifters---the pair functions as a second-order filter with an effective linewidth of approximately 900~MHz FWHM. An additional thermal phase shifter is embedded between each resonator pair to control the spectral phase of each channel. The resonance frequencies and spectral phases of each channel are measured using multi-heterodyne spectroscopy (MHS) and dual-comb spectroscopy (DCS), respectively~\cite{cohen2024silicon,wu2025chip}. A Python-based control routine tunes the resonance positions and phase profiles by driving the integrated thermo-optic phase shifters on the chip.

After configuring the pulse shaper through MHS and DHS, we align the phases of the EOPMs by tuning the continuous-wave (CW) laser to one of the side frequency bins ($\omega_{-1}$ or $\omega_2$), applying small modulation index, and adjusting Phase Shifters 2 and 3 to minimize the residual sideband power at the output. Note that a small modulation index is not necessary for this procedure, but simplifies the alignment by equating the out-of-phase condition  with minimal sideband power. 

To characterize gate performance, we follow coherent-state probing techniques \cite{lu2018electro,rahimi2011quantum,rahimi2013direct}. Initially, the transformation is probed by tuning the CW laser to $\omega_0$ or $\omega_1$, and the resulting output spectrum is recorded.  
Then, to extract the beamsmplitter phase, a dual-line input is prepared by an electro-optic intensity modulator (EOIM) biased at its null point for carrier suppression, and driven at half the mode-spacing frequency. This produces two main sidebands separated by the desired frequency bin spacing. The relative phase between the two spectral components is adjusted by introducing a delay in the radio-frequency (RF) modulation signal via Phase Shifter 1. For both single- and dual-line tests, spectral outputs are recorded using an optical spectrum analyzer (OSA) with 150~MHz resolution and 5~s integration time per spectrum. Each measurement is repeated five times to obtain error bars. By combining results from both tests, we can reconstruct the experimental transformation matrix  (Appendix \ref{appendixC}). For the tunable beamsplitter experiments, $\alpha$ or $\theta$ is detuned from the Hadamard gate setting while all other parameters remain fixed. The resulting spectral outputs for single-line input states are recorded to extract the corresponding splitting ratios.

\section{Results and discussion}
\subsection{Ultrafine resolution frequency-bin Hadamard gate}
\begin{figure}[tb!]
\centering\includegraphics[width=13.2cm]{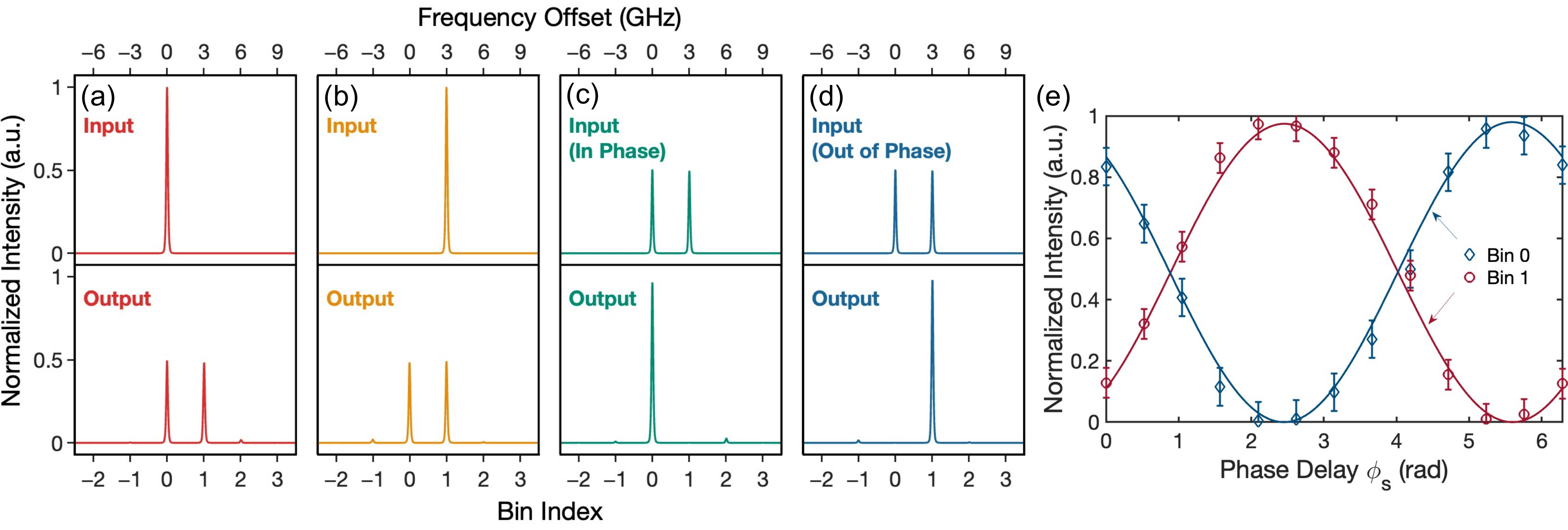}
\caption{Experimentally measured spectral outputs for the frequency-domain Hadamard gate with 3~GHz mode spacing. (a) Single-bin input $|0_{\omega_0}, \alpha_{\omega_1}\rangle$. (b) Single-bin input $|\alpha_{\omega_0}, 0_{\omega_1}\rangle$.
(c) In-phase two-bin input $|\alpha_{\omega_0},\alpha_{\omega_1}\rangle$. (d) Out-of-phase two-bin input $|\alpha_{\omega_0},-\alpha_{\omega_1}\rangle$.
(e) Relative output intensities in bins 0 and 1 for general two-bin input $|\alpha_{\omega_0}, e^{i(\phi_i + \phi_s)} \alpha_{\omega_1}\rangle$, as the adjustable phase $\phi_s$ is swept over $2\pi$.
}
\label{fig3}
\end{figure}

Figure \ref{fig3} presents the experimental results for the Hadamard gate with all six available pulse-shaper channels, i.e., the configuration $(B,\alpha,\theta)=(6,\pi,0.8283)$. 
The output spectra for single-line input states at bins 0 and 1 [Figs.~\ref{fig3}(a) and \ref{fig3}(b)] demonstrate nearly equal energy splitting between the computational basis modes. Small residual features at bin $-1$ and $2$ indicate nonunity of $\widetilde{\sP}$ as expected from theory. 
For dual-line characterization, we measure the output intensity distribution across bins 0 and 1 [Fig.~\ref{fig3}(e)] when adjusting Phase Shifter 1 (Fig.~\ref{fig2}). Denoting the phase delay introduced by Phase Shifter 1 as $\phi_s$, the total relative phase between the dual-line coherent-state input $|\alpha_{\omega_0}, e^{i\phi}\alpha_{\omega_1}\rangle$ is given by $\phi = \phi_i + \phi_s$, where $\phi_i$ accounts for any fixed phase offset due to inherent linear delay of our system. We define 
$\phi = 0$ when the output intensity in mode 0 is maximized. 
Figures \ref{fig3}(c) and \ref{fig3}(d) show the results for in-phase and out-of-phase dual-line inputs, respectively. These outputs closely resemble the time-reversed versions of the single-line testing shown in Figs. \ref{fig3}(a) and \ref{fig3}(b), consistent with the property $H^2 = I$. From these measurements, we obtain the fidelity $\sF = 0.99996(12)$, matching to within uncertainty the highest experimental frequency beamsplitter fidelity ever reported in bulk systems: $\sF = 0.99998(3)$ in \cite{lu2018electro}. Our measured modified success probability of $\widetilde{\sP} = 0.9768(23)$ also agrees closely with the theoretical prediction $\widetilde{\sP}=0.9747$.
Given the significantly more involved calibration procedures for the integrated pulse shaper  (e.g., MHS and DHS)  compared to bulk devices, such matching of tabletop QFP fidelities inspires an encouraging outlook for high-fidelity frequency-bin processing on chip.


One notable advantage of the QFP architecture is its inherent parallelizability \cite{lu2018electro}, allowing for simultaneous beamsplitter operations across multiple, spectrally distinct frequency bands. However, realizing such parallel operations in practice requires the pulse shaper to provide sufficient spectral channels to ensure both amplitude transmission and phase control across the relevant frequency bins. Increasing the number of channels in our microring-resonator-based pulse shaper introduces several engineering challenges. As the number of rings increases, the control complexity of configuring the spectral channels increases accordingly.

\begin{figure}[tb!]
\centering\includegraphics[width=8.8cm]{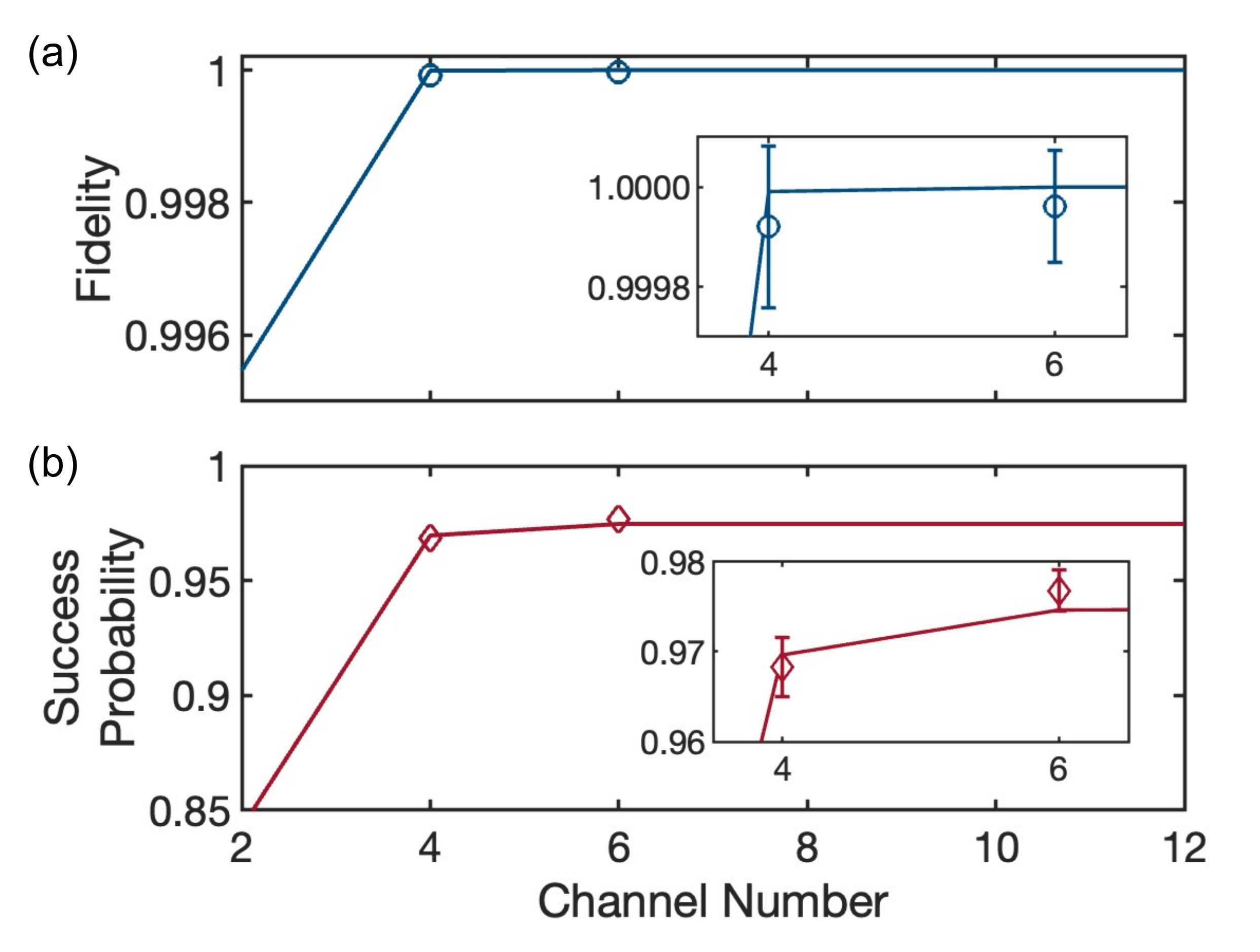}
\caption{(a) Fidelity $\sF$ and (b) success probability $\widetilde{\sP}$ as functions of the (even) number of spectral channels in the pulse shaper. Solid lines represent theoretical curves, while experimental data are shown as markers. Insets show enlarged views of the data. Note that the error bars indicate standard deviations and do not imply that the measured $\sF$ exceeds the theoretical upper bound of unity.}
\label{fig4}
\end{figure}

Additionally, for thermally tuned devices, heat dissipation becomes increasingly problematic. 
Thermal crosstalk between heaters can degrade tuning accuracy and significantly prolong the convergence time for reaching target resonances and phase profiles. Therefore, it is essential to use each channel as efficiently as possible and to minimize the number of required channels without compromising gate performance. 
To evaluate this tradeoff, we depict the theoretical Hadamard gate performance as a function of the number of pulse-shaper channels $B$ in \cref{eq:Vbs} as lines in Fig.~\ref{fig4}. 
The results show that performance saturates at $B=6$ channels but with minimal reduction at $B=4$, $\widetilde{\sP}$ reducing by merely $\approx$1\%. This behavior can be understood by noting that 
the relative power in the second-order sidebands (lost in the case of only four bins) after the first EOPM is $|J_2(\theta)/J_0(\theta)|^2 \approx 0.009$ at our modulation index $\theta = 0.8283$. For higher orders (e.g., $J_3$), the contribution becomes even smaller---thus the majority of spectral content remains confined within the four-bin window.

To experimentally validate the performance with $B=4$, we reprogram the pulse shaper to use only four channels with a spectral phase pattern $(0, 0, \pi, \pi)$, while leaving the remaining two channels detuned from the frequency-bin grid. 
The resulting gate achieves $\sF = 0.99992(16)$ and $\widetilde{\sP}=0.9683(33)$, in close agreement with theory $(\sF,\widetilde{\sP}) = (0.999991,0.9696)$, as shown by the discrete points in \cref{fig4} for both the $B=4$ and previous $B=6$ implementation.
While the $\lesssim$1\% reduction in $\widetilde{\sP}$ may accumulate for certain quantum information processing protocols involving many sequential gates, we confirm that four channels are sufficient for performing a single beamsplitter operation within a given frequency band, supporting the viability of resource-efficient designs for QFP-based platforms. Indeed, from a SWAP perspective, we note that the $B=6$ heater settings draw 474.02~mW of electrical power in our particular experiment, while the $B=4$ beamsplitter only requires 320.88~mW---highlighting wall-plug energy savings above and beyond questions of spectral efficiency alone.

\begin{figure}[tb!]
\centering\includegraphics[width=9.2cm]{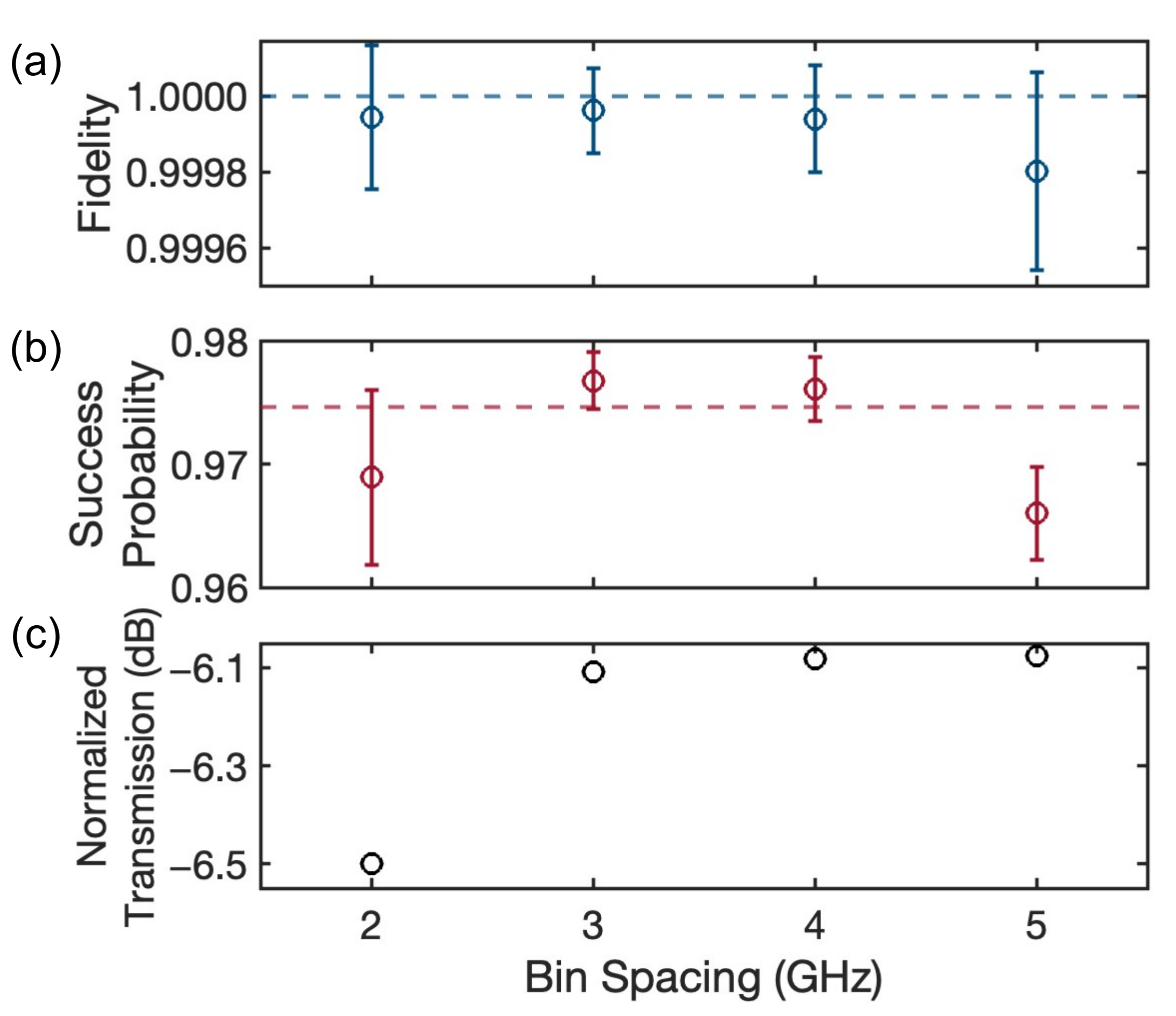}
\caption{Gate performance as a function of frequency spacing. Experimentally measured (a) fidelity $\sF$ and (b) success probability $\widetilde{\sP}$ overlaid with corresponding theoretical values (dashed lines). Error bars reflect measurement uncertainty. (c) Average optical transmission across the six spectral channels, normalized by the $\sim$8.5 dB fiber-to-chip coupling loss. 
}
\label{fig5}
\end{figure}

Next, we demonstrate the tunability of the frequency-bin spacing in our system. While \cite{cohen2024silicon} has previously shown the ability to tune the channel spacing $\Delta f$ between 3--5 GHz for this device, we find that the lower limit can be extended to 2 GHz. As shown in Fig.~\ref{fig5}, the gate performance remains consistent across the 2--5 GHz range, with  $\sF>0.9995$ and $\widetilde{\sP}>0.9621$. 
One observed drawback of operating at a 2 GHz frequency spacing is a reduction in overall optical transmission [\cref{fig5}(c)]. 
This degradation is likely due to interchannel crosstalk \cite{nussbaum2022design}: as the spectral separation between cavity resonances decreases, partial coupling between adjacent resonators can occur, introducing additional loss. Attempts to operate below 2~GHz were unsuccessful, primarily due to the 250~MHz resolution limit of our MHS system used for resonance probing \cite{cohen2024silicon,wu2025chip}. Given the 1.3~GHz individual ring linewidth, such narrow spacings (e.g., $\Delta f \approx$  1 GHz) cause even small variations in tuning current to overlap adjacent resonances, 
ultimately forcing the control algorithm to fail. Nevertheless, it is worth emphasizing that the narrowest QFP operation previously reported employed an 18 GHz bin spacing \cite{lu2018electro}. Realizing a high-fidelity Hadamard gate at 2 GHz therefore represents nearly an order-of-magnitude advance in resolution.

\subsection{Arbitrary beamsplitters}

Beamsplitters with arbitrary splitting ratios are fundamental tools in quantum information science. They allow coherent control over quantum states across the full Bloch sphere \cite{lu2020fully}, underpin linear-optical quantum circuits based on the Reck decomposition~\cite{Reck1994, Clements2016}, and form critical building blocks in protocols such as boson sampling~\cite{aaronson2010computationalcomplexitylinearoptics}. 
In the following sections, we present two distinct methods for implementing frequency-domain beamsplitters with continuously tunable splitting ratios at a mode spacing of 3~GHz.

\begin{figure}[b!]
\centering\includegraphics[width=13cm]{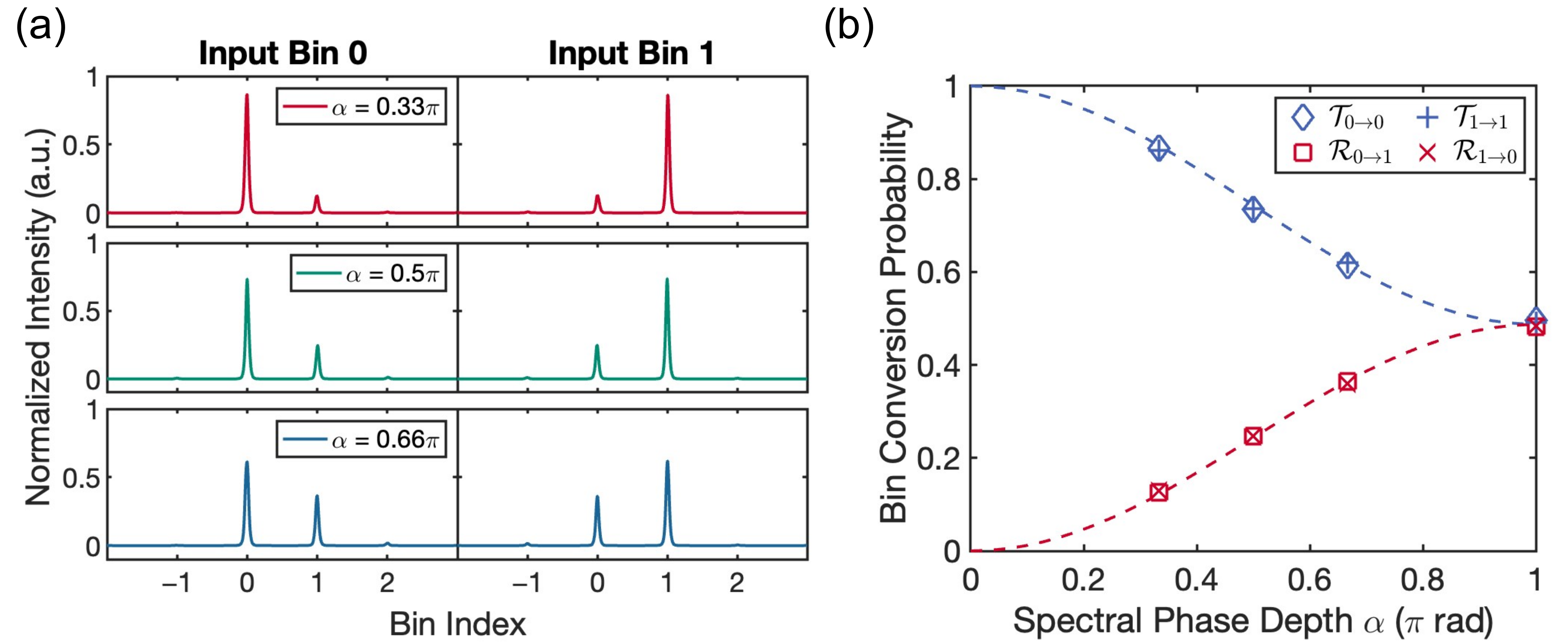}
\caption{Frequency beamsplitters with tunable splitting ratios via spectral phase control.
(a) Output spectra for single-frequency inputs at bin 0 (left) and bin 1 (right) for three different values of the spectral phase parameter $\alpha$.
(b) Extracted mode transmissivity ($\sT$) and reflectivity ($\sR$), compared with theoretical predictions (upper/blue dashed line for $\sT_{0\rightarrow 0}$ and $\sT_{1\rightarrow 1}$, lower/red dashed line for $\sR_{0\rightarrow 1}$ and $\sR_{1\rightarrow 0}$). Error bars are comparable in size to the markers.
}
\label{fig6}
\end{figure}

As first introduced in \cite{lu2018quantum}, the reflectivities $\sR$ and transmissivities $\sT$ of the frequency beamsplitter [\cref{eq:RandT}] can be tuned by varying the spectral phase parameter $\alpha$ in the pulse shaper. In addition to the Hadamard configuration at $\alpha = \pi$, we experimentally demonstrate three representative cases corresponding to $\alpha \in\{ \pi/3, \pi/2, 2\pi/3\}$ as shown in Fig.~\ref{fig6}(a). The output spectra, obtained by sending single-frequency inputs at both modes 0 and 1, clearly reveal distinct splitting ratios. A quantitative comparison between measured values of $\sR$ and $\sT$ and the theoretical predictions is provided in Fig.~\ref{fig6}(b).

\begin{figure}[tb!]
\centering\includegraphics[width=8cm]{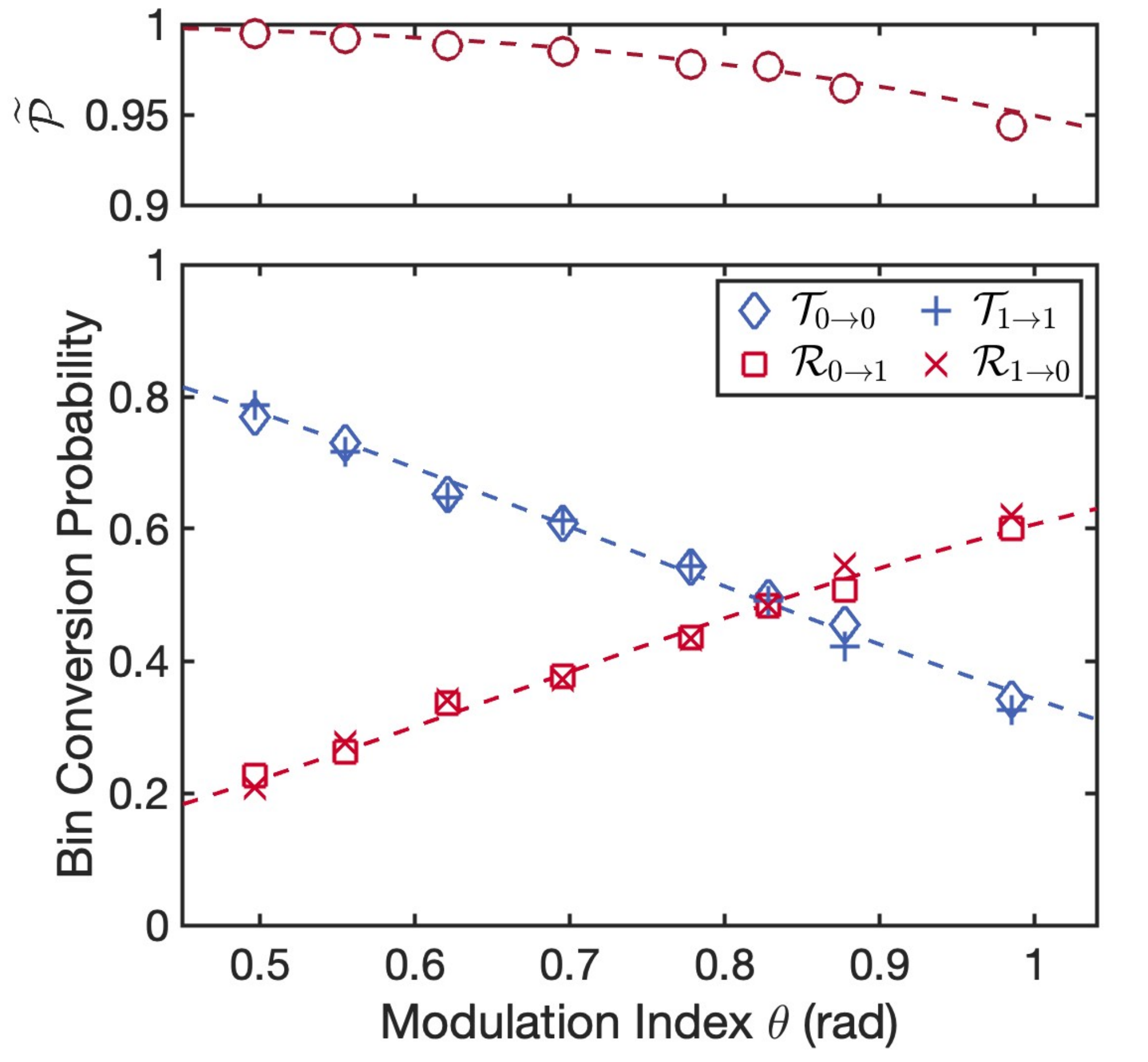}
\caption{Tunable beamsplitter implemented through adjusting the modulation index. Top: success probability $\widetilde{\sP}$; bottom: mode transmissivity ($\sT$) and reflectivity ($\sR$) as functions of the modulation index $\theta$. Dashed lines represent theoretical predictions, while markers denote experimental data. Error bars are comparable to the markers in size. 
}
\label{fig7}
\end{figure}

Although continuous variation of the splitting ratio can be achieved by smoothly tuning $\alpha$, reconfiguring our pulse shaper chip is relatively time-consuming (on the order of seconds). 
Accordingly, to enable dynamic control over the splitting ratio, we introduce a novel alternative approach by tuning the modulation index $\theta$ through adjustments to the RF drive power applied to the EOPMs---a quantity that in principle can be adjusted on nanosecond timescales. 
Experimental results for $\theta \in [0.5,1.0]$ are shown in Fig.~\ref{fig7}. For $\theta < 0.8283$, the splitting ratio evolves in a fashion qualitatively similar to the phase-controlled results in Fig.~\ref{fig6}. However, when $\theta > 0.8283$, the system behaves as a frequency shifter, with mode-hopping probabilities ($\sR_{0\rightarrow 1}$ and $\sR_{1\rightarrow 0}$) exceeding the mode-preserving probabilities ($\sT_{0\rightarrow 0}$ and $\sT_{1\rightarrow 1}$). 
As highlighted by by the measured success probabilities, the regime $\sR>\sT$ is marked by a steady reduction in $\widetilde{\sP}$. Indeed, previous work~\cite{lu2020fully} has shown that a three-element QFP with single-tone RF modulation can at best reach $\sP=0.7590$ in the full shift limit $\sT\rightarrow 0$ (a Pauli $X$ gate), indicating further reductions in $\widetilde{\sP}$ are expected for additional $\theta$ tuning.
Consequently, we did not explore values beyond $\theta > 1.0$, since excessive sideband generation outside the six-channel window renders the output less useful for practical applications. Nonetheless, whenever the observed reduction in $\widetilde{\sP}$ can be tolerated, this approach offers a simple pathway for rapid frequency beamsplitter adjustments beyond 50/50 split ratios. 

\section{Conclusion}
In conclusion, we have demonstrated high-performance frequency beamsplitters enabled by an on-chip pulse shaper. The gate retains excellent fidelity even when operating with only four spectral channels, highlighting the platform's spectral scalability in addition to its compact footprint. The ultranarrow channel linewidths support tunable frequency spacings down to 2 GHz---an order-of-magnitude improvement over previous QFP implementations. Arbitrary and continuously tunable splitting ratios are achieved via control over either spectral phase or modulation index. 
The ability to implement ideal frequency beamsplitters enables advanced experiments involving quantum interference and correlation control with nonclassical light in the spectral domain \cite{lu2018quantum,imany201850}. 
Furthermore, the demonstrated tuning range of 2--5~GHz suggests intriguing paths for quantum transduction between optical and RF domains. Many solid-state quantum platforms~\cite{doherty2013nitrogen,kjaergaard2020superconducting,krantz2019quantum}, such as diamond nitrogen-vacancy centers (2.87 GHz) and superconducting transmon qubits (3--6 GHz), couple naturally to RF or microwave photons, whereas optical photons are far more suitable for long-distance quantum communication. Bridging this spectral gap requires efficient RF-to-optical frequency conversion \cite{awschalom2021development}. 
While the commercial EOPMs employed in the current QFP do not offer microwave-to-optical conversion efficiencies sufficient for reasonable transduction at single-microwave-photon levels, our pulse shaper's ability to control optical frequency-bin states with few-GHz spacings suggests exciting opportunities for future integration with quantum transduction technologies tailored to such RF qubits.

Currently, the total insertion loss of the QFP is approximately 21 dB, comprising $\sim$15~dB from the on-chip pulse shaper ($\sim$6.5~dB drop-port loss plus $\sim$8.5~dB fiber-to-chip coupling loss) and $\sim$3~dB per EOPM (off-chip). Although similar losses have been tolerated in prior quantum experiments\cite{wu2025chip}, further integration, such as incorporating the EOPMs directly on chip, could reduce the total system loss to levels lower than fully tabletop implementations \cite{nussbaum2022design}. In addition, realizing high-dimensional quantum gates \cite{lu2022high} or parallel beamsplitter operations will require pulse shapers with a greater number of spectral channels. These considerations underscore the importance of developing more advanced and densely integrated photonic platforms. 
With continuing progress in integrated photonic technology~\cite{myilswamy2025chip}, realizing a fully integrated QFP system is promising. The developments of high-speed silicon-organic hybrid modulators \cite{alloatti2014100} and heterogeneously integrated thin-film lithium niobate on silicon modulators \cite{he2019high} provide excellent platforms to integrate the phase modulators. Silicon-based on-chip quantum light sources have been widely demonstrated \cite{clementi2023programmable} and are therefore ready to incorporate into QFP circuits. 

\section*{Appendices}
\appendix
\setcounter{equation}{0}\renewcommand{\theequation}{\thesection.\arabic{equation}}
\setcounter{figure}{0}\renewcommand{\thefigure}
{\thesection\arabic{figure}}

\section{QFP mode transformation}
\label{appendixA}
As discussed in the main text, the QFP can be modeled as a mode transformation $V_{mn}$ between the input and output annihilation operators $\hat{a}_n$ and $\hat{b}_m$. In general, for microring-based pulse shapers, the Lorentzian filter profiles produce appreciable intrabin amplitude and phase effects that require modeling the full QFP transformation as a function of frequency offset from each bin center; a full model accounting for these effects can be found in  \cite{nussbaum2022design}. However, because the experiments here rely on a periodically phase-modulated CW laser whose linewidth ($\sim$100~kHz) is far narrower than the filter FWHM, these intrabin effects can be neglected and the full frequency transformation written the fully discrete form~\cite{lu2020fully}:
\begin{equation}
\label{eq:V}
\hat{b}_m = \sum_{n=0}^{N-1} V_{mn} \hat{a}_n = \sum_{n=0}^{N-1} \left(\sum_{k=-\infty}^{\infty} d_{m-k}T_k c_{k-n} \right) \hat{a}_n,
\end{equation}
where $N$ is the number of bins in the computational space. The last equality holds for a three-element QFP (EOPM--pulse shaper--EOPM), with a pulse-shaper filter $T_k$ and mode-hopping coefficients $c_{k-n}$ and $d_{m-k}$ for the first and second EOPMs, respectively. Note that $T$, $c$, and $d$ are, in principle, matrices, but the spectral filter $T$ is diagonal in the frequency domain and can therefore be reduced to a single-index quantity. On the other hand, the mode-hopping operators $c$ and $d$ exhibit Toeplitz structure as a consequence of the translation symmetry of phase modulation.

For the operations examined in this work, the EOPMs are driven by out-of-phase sinusoidal signals with modulation index $\theta$, leading to temporal phase factors of the form $e^{\pm i\theta\sin\Delta \omega t}$. Using the Jacobi--Anger expansion,
\begin{equation}
e^{\pm i\theta\sin(\Delta \omega t)} = \sum_{k=-\infty}^{\infty} (\pm 1)^kJ_k(\theta) e^{i k \Delta \omega t},
\end{equation}
the mode-hopping coefficients reduce to Bessel functions of the first kind, namely, $d_\ell = (-1)^\ell c_\ell = J_\ell(\theta)$. Applying the identity $J_{-\ell}(\theta)=(-1)^\ell J_\ell(\theta)$, we obtain
\begin{equation}
\label{eq:V_bessel}
    V_{mn} = \sum_k J_{m-k}(\theta)\,T_k\,J_{n-k}(\theta).
\end{equation}
The pulse shaper contains an even number of active bins $B$ with an $\alpha$ phase shift between $k=0$ and $k=1$. Assuming uniform amplitude transmission across all channels and perfect out-of-band rejection, the transmission coefficients $T_k$ are given by
\begin{equation}
\label{eq:T}
T_k = \begin{cases}
1 & ; -\frac{B}{2}+1\leq k\leq0 \\
e^{i\alpha} & ; \phantom{-\frac{B}{2}+\;\,}1\leq k \leq \frac{B}{2} \\
0 & ; \quad\qquad\text{otherwise}
\end{cases}.
\end{equation}
Substituting Eq.~\eqref{eq:T} into Eq.~\eqref{eq:V_bessel} and rearranging the summation indices, we obtain the full expression for the mode transformation as given by \cref{eq:Vbs} in the main text.

\section{Modified success probability}
\label{appendixB}
Due to the translation symmetry of electro-optic modulation in the frequency domain, beamsplitters based on EOPMs are generally nondeterministic, allowing output photons to scatter outside the computational bins. To quantify the probability that photons remain in the bins of interest, the success probability was originally defined as \cite{lukens2016frequency,lu2022high}:
\begin{equation}
\label{eq:P}
\sP  = \frac{\Tr W^\dagger W}{\Tr U^\dagger U} = \frac{1}{N}\Tr W^\dagger W.
\end{equation}
Although this is well defined \emph{theoretically} for any QFP of the form in \cref{eq:V}, \emph{experimental} confirmation of \cref{eq:P} presents unique subtleties for the on-chip shaper when compared to bulk versions. In particular,  previous QFP experiments have leveraged power conservation of the full transformation, i.e.,
\begin{equation}
\label{eq:normOld}
\sum_{m=-\infty}^\infty\sum_{n=0}^{N-1} |V_{mn}|^2 = N,
\end{equation}
to back out insertion losses and normalize the empirically obtained coefficients $V_{mn}$~\cite{lu2018electro,Lu2019a,lu2020fully}. Essentially, because the lossless case requires all input light to leave in \emph{some} output bin $m$, one need only measure the power in all output bins for each input computational state to evaluate fundamental gate performance apart from technical sources of insertion loss. Yet as seen in the bottom of \cref{fig1}(b), in the on-chip pulse shaper, frequency bins that are not explicitly downloaded by a ring filter are irretrievably lost; they exit the through port and never reach the QFP output. Accordingly, at the output it is no longer possible to distinguish between insertion loss and gate loss: how many photons fail to reach the output due to technical losses versus how many photons are lost by the design of the finite-bandwidth gate itself? This distinction means that the quantity obtained by normalizing to the output power is not success probability \emph{per se}, but rather a ``modified success probability''
\begin{equation}
\label{eq:modP}
\widetilde{\sP}  = \frac{\Tr W^\dagger W}{\Tr V^\dagger V} = \frac{1}{\eta N}\Tr W^\dagger W,
\end{equation}
where $\eta$ is a positive number no greater than unity because $\Tr V^\dagger V\leq N$. The inequality  $\sP\leq\widetilde{\sP}\leq 1$ holds in general, 
where $\widetilde{\sP}=\sP$ is obtained in the limit of many pulse-shaper channels $B$, again assuming these channels follow the definition in \cref{eq:T}: 
\begin{equation}
\label{eq:limit}
\lim_{B\rightarrow\infty}\widetilde{\sP}  = \sP.
\end{equation}

For most cases considered in this work (e.g. $B=6$ and $\theta \le 1$), theoretical evaluation of Eqs.~(\ref{eq:P}) and (\ref{eq:modP}) shows that the difference between $\sP$ and $\widetilde{\sP}$ is negligible ($\sim$10$^{-4}$), as the majority of the sideband power remains within the six-channel spectral window. In contrast, for the $B=4$ case, $\widetilde{\sP}$ exceeds $\sP$ by approximately $7\times10^{-3}$. Despite the small numerical difference between the two definitions, we stick with the modified success probability to account for subtle differences between on-chip pulse shapers, which typically have narrower passbands, and bulk devices, for which large-bandwidth implementations are commercially available. 
Moreover, given the practical difficulty of scaling the number of channels in on-chip pulse shapers, high-dimensional gates cannot always be implemented by arbitrarily increasing the available spectral channels. 
In such cases, multitone phase modulation is more likely to scatter optical power into sidebands outside the passband, and the modified success probability offers a more suitable metric.

\section{Estimation of gate fidelity and success probability}
\label{appendixC}
The validity of probing quantum optical multiport operations using coherent states has been established in~\cite{rahimi2011quantum,rahimi2013direct}. In this work, we adopt the notation and procedure outlined in \cite{lu2018electro} to reconstruct the transformation matrix elements for Hadamard gate experiments. Given that beamsplitting of coherent states produces equivalent outcomes under both quantum and classical treatments \cite{gerry2023introductory}, and is consistent with classical optics, a classical field approach suffices to model the essential system behavior. Under this framework, a general multi-frequency-bin optical field can be expressed as
\begin{equation}
    E(t) = \sum_m E_m e^{-i\omega_m t},
\end{equation}
where $E_m $ denotes the complex amplitude of the $m$-th bin. The input and output mode amplitudes are related by the linear transformation:
\begin{equation}
E_n^{\text{(out)}} = \sum_m V_{nm} E_m^{\text{(in)}},
\end{equation}
with the transformation matrix expressed in polar form as $V_{nm} = r_{nm} e^{i\phi_{nm}}$. Without loss of generality, the phases in the first column and row of $V_{nm}$ can be set to zero in recognition of the freedom to define phase reference planes at both the output and the input to the multiport~\cite{rahimi2013direct,lu2018electro,peres1989construction}. 

For unit-power input (i.e. $\sum_m |E_m^{\text{(in)}}|^2 = 1$), single-line tests provide access to the amplitude of the unknown $2\times 2$ transformation $W$ (recall that $W_{mn}= V_{mn}$ for $m,n\in\{0,1\}$) according to the following~\cite{lu2018electro}:
\begin{equation}
\begin{aligned}
L\begin{pmatrix}
r_{00} & r_{01} \\
r_{10} & r_{11} e^{i\phi_{11}}
\end{pmatrix}
\begin{pmatrix}
1 \\
0
\end{pmatrix}
&=L
\begin{pmatrix}
r_{00} \\
r_{10}
\end{pmatrix}
\xrightarrow{\text{OSA}}L^2
\begin{pmatrix}
r_{00}^2 \\
r_{10}^2
\end{pmatrix}
\\[1em]
L\begin{pmatrix}
r_{00} & r_{01} \\
r_{10} & r_{11} e^{i\phi_{11}}
\end{pmatrix}
\begin{pmatrix}
0 \\
1
\end{pmatrix}
&=L
\begin{pmatrix}
r_{01} \\
r_{11}e^{i\phi_{11}}
\end{pmatrix}
\xrightarrow{\text{OSA}}L^2
\begin{pmatrix}
r_{01}^2 \\
r_{11}^2
\end{pmatrix}.
\end{aligned}
\end{equation}
The extra prefactor $L$ before $W$ accounts for frequency-independent losses in the practical QFP system, including the insertion loss of the EOPMs, fiber-to-chip coupling loss, and the drop-port loss of the second-order filters. To focus on the intrinsic gate performance, we normalize each output spectrum by its total detected power, enforcing $\sum_n | E_n^{\text{(out)}}|^2 = 1$. This means that we obtain quantities scaled by the unmeasured parameter $\eta$ introduced in \cref{eq:modP}:
\begin{equation}
\begin{aligned}
L^2\begin{pmatrix}
r_{00}^2 \\
r_{10}^2
\end{pmatrix}
&\xrightarrow{\text{Norm.}}
\frac{1}{\eta}
\begin{pmatrix}
r_{00}^2 \\
r_{10}^2
\end{pmatrix}
\equiv
\begin{pmatrix}
\gamma_{00}^2 \\
\gamma_{10}^2
\end{pmatrix}
\\[1em]
L^2\begin{pmatrix}
r_{01}^2 \\
r_{11}^2
\end{pmatrix}
&\xrightarrow{\text{Norm.}}
\frac{1}{\eta}
\begin{pmatrix}
r_{01}^2 \\
r_{11}^2
\end{pmatrix}
\equiv
\begin{pmatrix}
\gamma_{01}^2 \\
\gamma_{11}^2
\end{pmatrix}
\end{aligned}
\end{equation}
Note that this normalization also reduces uncertainty between repeated measurements arising from output power fluctuations caused by environmental thermal drift of the chip resonances.

In contrast to previous QFP error propagation approaches that computed $\sF$ and $\sP$ for separate trials to aid in uncertainty quantification~\cite{lu2018electro}, here we exploit direct error propagation of the measured optical powers. By repeating these measurements five times, we obtain the sample means $\langle \cdot \rangle$, variances $\text{Var}(\cdot)$, and covariances $\text{Cov}(\cdot, \cdot)$ of the quantities $\gamma_{nm}$. The modified success probability can be calculated through \cref{eq:modP}:
\begin{equation}
    \widetilde{\sP} = \frac{\Tr W^\dagger W}{\Tr V^\dagger V} = \frac{\gamma^2_{00} +\gamma^2_{01} + \gamma^2_{10} + \gamma^2_{11}}{2} \equiv \widetilde{\sP}(\gamma_{00}, \gamma_{01}, \gamma_{10}, \gamma_{11})
\end{equation}
We estimate the measured value as 
\begin{equation}
    \widetilde{\sP}_{\text{meas}} = \widetilde{\sP}(\langle \gamma_{00} \rangle, \langle \gamma_{01} \rangle, \langle \gamma_{10} \rangle, \langle \gamma_{11} \rangle) \pm \Delta \widetilde{\sP},
\end{equation}
with  $\Delta \widetilde{\sP}$ calculated by error propagation:
\begin{equation}
(\Delta \widetilde{\sP})^2 = \sum_{\substack{m,n \\m',n'}\in\{0,1\}} \frac{\partial\widetilde{\sP}}{\partial \gamma_{mn}}  \frac{\partial\widetilde{\sP}}{\partial \gamma_{m'n'}} \text{Cov}(\gamma_{mn}, \gamma_{m'n'}).
\end{equation}
From dual-line testing, we can analyze the transformation
\begin{equation}
L\begin{pmatrix}
r_{00} & r_{01} \\
r_{10} & r_{11} e^{i\phi_{11}}
\end{pmatrix}
\cdot
\frac{1}{\sqrt{2}}
\begin{pmatrix}
1 \\
e^{i(\phi_i + \phi_s)}
\end{pmatrix}
=
\frac{L}{\sqrt{2}}
\begin{pmatrix}
r_{00} + r_{01} e^{i(\phi_i + \phi_s)} \\
r_{10} + r_{11} e^{i(\phi_i + \phi_s + \phi_{11})}
\end{pmatrix},
\end{equation}
such that the measured (normalized) optical power from each mode using the OSA is given by
\begin{equation}
\frac{1}{2}
\begin{pmatrix}
\gamma_{00}^2 + \gamma_{01}^2 + 2\gamma_{00}\gamma_{01} \cos(\phi_i + \phi_s) \\
\gamma_{10}^2 + \gamma_{11}^2 + 2\gamma_{10}\gamma_{11} \cos(\phi_i + \phi_s + \phi_{11})
\end{pmatrix}
\equiv
\begin{pmatrix}
\rho_0 \\
\rho_1
\label{A.8}
\end{pmatrix}.
\end{equation}
Here, $\phi_i$ represents the inherent linear delay in the system, and $\phi_s$ is the tunable phase introduced by the RF delay line. By scanning $\phi_s$ over a $2\pi$ range [as shown in Fig. \ref{fig3}(e)], and substituting the mean values $\langle \gamma_{nm} \rangle$ obtained from single-line measurements into \cref{A.8}, we fit $\rho_0$ to extract $\phi_i$, and then fit $\rho_1$ to extract $\phi_{11}$. The variance $\text{Var}(\phi_{11})$ is estimated by squaring one-fourth of the 95$\%$ confidence interval width of the fitted $\langle\phi_{11}\rangle$. The fidelity is calculated as (with $U$ the Hadamard gate):
\begin{equation}
\sF = \frac{\Tr U^\dagger W \Tr W^\dagger U}{\Tr U^\dagger U \Tr W^\dagger W} 
= \frac{|\gamma_{00} + \gamma_{01} + \gamma_{10} - \gamma_{11} e^{i\phi_{11}}|^2}{4(\gamma_{00}^2 + \gamma_{01}^2 + \gamma_{10}^2 + \gamma_{11}^2)}
\equiv \sF(\gamma_{00}, \gamma_{01}, \gamma_{10}, \gamma_{11}, \phi_{11})
\end{equation}
The measured value is estimated by:
\begin{equation}
    \sF_{\text{meas}} = \sF(\langle \gamma_{00} \rangle, \langle \gamma_{01} \rangle, \langle \gamma_{10} \rangle, \langle \gamma_{11} \rangle, \langle \phi_{11} \rangle) \pm \Delta \sF
\end{equation}
where the uncertainty $\Delta \sF$ is again calculated using error propagation:
\begin{equation}
(\Delta \sF)^2 = \sum_{\substack{m,n \\m',n'}\in\{0,1\}} \frac{\partial \sF}{\partial \gamma_{mn}}  \frac{\partial \sF}{\partial \gamma_{m'n'}} \text{Cov}(\gamma_{mn}, \gamma_{m'n'})+
 \left( \frac{\partial \sF}{\partial \phi_{11}} \right)^2 \text{Var}(\phi_{11}).
\end{equation}  

\begin{backmatter}
\bmsection{Funding}
Funding was provided by the National Science Foundation (2034019-ECCS) and the U.S. Department of Energy (CW31444).

\bmsection{Acknowledgment}
The authors thank Matthew van Niekerk and Stefan Preble for providing wirebonding materials. Preliminary results of this paper were presented at CLEO 2025 as paper number FF118\_1.

\bmsection{Disclosures}
The authors declare no conflicts of interest.

\bmsection{Data Availability}
Data available from the authors upon reasonable request.

\end{backmatter}


\bibliography{sample}






\end{document}